\documentclass{llncs}  

\usepackage{graphicx} 

\usepackage[T1]{fontenc}
\usepackage{amsmath}
\usepackage[english]{babel}
\usepackage{array,tabularx}
\usepackage{booktabs} 
\usepackage{makeidx}  % allows for index generation
\usepackage{xparse}
\usepackage{moredefs}
\usepackage{lips}
\usepackage{epstopdf} 
\usepackage{color}
\usepackage{csquotes}
\usepackage{xcolor}
\usepackage{times}
\usepackage[hidelinks]{hyperref}
\usepackage[nolist,nohyperlinks]{acronym}
\usepackage{comment}
\usepackage{paralist}
\usepackage{cleveref}
\usepackage{wrapfig}
\usepackage{multirow}
\usepackage{microtype}
\usepackage{listings}
\usepackage[super]{nth}
\usepackage{natbib}

\usepackage{float}
\floatstyle{plaintop}
\restylefloat{table}

\usepackage[style=base,labelfont=bf,labelsep=period,tableposition=top,font=small]{caption}
\makeatletter
\renewcommand\@biblabel[1]{#1.}
\makeatother
\addto\captionsenglish{}

\usepackage{fancyhdr}
\fancypagestyle{WI_footer}{\fancyhf{}\fancyfoot[L]{\color{black}\scriptsize AI Act Workshop, 19th International Conference on Wirtschaftsinformatik\\September 2024, Würzburg, Germany \\\textbf{Accepted Manuscript}}}

\crefformat{footnote}{#2\footnotemark[#1]#3}

\expandafter\def\expandafter\UrlBreaks\expandafter{\UrlBreaks%  save the current one
  \do\a\do\b\do\c\do\d\do\e\do\f\do\g\do\h\do\i\do\j%
  \do\k\do\l\do\m\do\n\do\o\do\p\do\q\do\r\do\s\do\t%
  \do\u\do\v\do\w\do\x\do\y\do\z\do\A\do\B\do\C\do\D%
  \do\E\do\F\do\G\do\H\do\I\do\J\do\K\do\L\do\M\do\N%
  \do\O\do\P\do\Q\do\R\do\S\do\T\do\U\do\V\do\W\do\X%
  \do\Y\do\Z}

\newcolumntype{L}[1]{>{\raggedright\arraybackslash}p{#1}}   % linksbündig mit Breitenangabe
\newcolumntype{C}[1]{>{\centering\arraybackslash}p{#1}}     % zentriert mit Breitenangabe
\newcolumntype{R}[1]{>{\raggedleft\arraybackslash}p{#1}}    % rechtsbündig mit Breitenangabe

\begin{document}
\frontmatter          % for the preliminaries

\mainmatter              % start of the contributions

\title{Towards Assuring EU AI Act Compliance \\ and Adversarial Robustness of LLMs}

%\author{Blinded\inst{1}}
%\authorrunning{Blinded et al.} % abbreviated author list (for running head)
%\institute{Blinded}
\subtitle{Research in Progress} %Specify type of research paper here!
\author{Tomas Bueno Momcilovic\inst{1} \and
Beat Buesser\inst{2} \and
Giulio Zizzo\inst{3} \and \\
Mark Purcell\inst{3} \and 
Dian Balta\inst{1}
}

\institute{
fortiss GmbH Research Institute, Munich, Germany\\
%\email{\{momcilovic,balta\}@fortiss.org} 
\and
IBM Research Europe, Zurich, Switzerland\\
%\email{beat.buesser@ibm.com} 
\and
IBM Research Europe, Dublin, Ireland\\
%\email{giulio.zizzo2@ibm.com, markpurcell@ie.ibm.com}
}

% -----------------------
% |  Begin of Document  |
% -----------------------
\maketitle
\setcounter{footnote}{0}

% ------------- 
% |  Abstract and Keywords  |
% -------------
\begin{abstract}
Large language models are prone to misuse and vulnerable to security threats, raising significant safety and security concerns. The European Union's Artificial Intelligence Act seeks to enforce AI robustness in certain contexts, but faces implementation challenges due to the lack of standards, complexity of LLMs and emerging security vulnerabilities. Our research introduces a framework using ontologies, assurance cases, and factsheets to support engineers and stakeholders in understanding and documenting AI system compliance and security regarding adversarial robustness. This approach aims to ensure that LLMs adhere to regulatory standards and are equipped to counter potential threats.\\

\end{abstract}

\thispagestyle{WI_footer}

% ------------- 
% |  Content  |
% -------------

\section{Introduction}

Large language models (LLMs) have shown great results in generating content from the data they were trained or fine-tuned on, when prompted in natural language \citep{kojima2022}. However, recent work shows that training, fine-tuning, prompting and generating can be vulnerable to malicious or accidental misuse \citep{yao2024llmsecurity}, as the models themselves are brittle to adversarial attacks \citep{Zou2023_Universal}. By exploiting unknown properties of LLMs, attacks can negatively impact the privacy and fundamental rights of EU citizens by leaking information or generating toxic content \citep{gdpr2016}. In combination with advanced capabilities (e.g., robotic control; \citealp{vemprala2023chatgpt}), applications (e.g., autonomous decision-making; \citealp{wang2024survey}) or contexts (e.g., medical diagnosis; \citealp{thirunavukarasu2023medicine}), compromised LLMs can also have safety implications.

The recently adopted EU Artificial Intelligence Act (further: EUAIA; \citealp{euaia2024corrigendum}) aims to mitigate the negative impact of "high-risk" AI systems by imposing demands on providers and deployers in designated contexts. Two foreseeable issues will make implementation of the Act considerably challenging if such systems have LLM components. First, the standards that operationalize the legal language into technical requirements are yet to be established, and rapid pace of development could render some parts obsolete. Second, the architecture of an LLM is substantially more dynamic, opaque and extensible than that of many predecessor models. Their performance, generality and trainable "harmlessness" are relatively novel breakthroughs that are not yet well-understood and brittle to even small changes. Thus, ensuring security with stable, proactive and mature practices is difficult.

In this work-in-progress, we investigate the problem and potential resolution for fulfilling the LLM- and robustness-relevant duties in EUAIA. We argue that to have justifiable confidence that LLMs are compliant and trustworthy, engineers need to continuously integrate, monitor, patch and communicate about the implemented defenses against adversarial attacks. We introduce a framework for knowledge representation and reasoning about the provenance, necessity and sufficiency of demands and defenses. Using ontologies, assurance cases and factsheets, the framework is intended to assist engineers and legal stakeholders in establishing a complete and dynamic picture of the safety, security and compliance of the LLM.

\section{Background}

The EUAIA \citep{euaia2024corrigendum} is a law covering particular aspects of AI usage in the EU, which was proposed in April 2021 and adopted in March 2024\footnote{The original text \citep{euaia2021original} has been drafted before the breakthrough of conversational LLMs in 2022 (e.g., ChatGPT; \citealp{openai2022}), and subsequently revised to include stipulations for LLM-like models \citep{euaia2024corrigendum}. We refer to the revised version that is made available by the \cite{fli2024euaia}.}. It is expected to enter into force in 2026, whereby technical standards and guidelines that interpret the Act will be available at the earliest in mid-2025 \citep{cen2024standards}, or no later than 2028 (Art. 6 Para. 5; Art. 15 Para. 2; EUAIA).

The core of EUAIA are duties placed upon the providers\footnote{i.e., those who develop or commission it, and put it on the market or into service; Art. 3, EUAIA.} of any AI system that will be used in high-risk domains (Art. 6-49; Annex I Section B; Annex III) or within regulated products (Annex I Section A). Other duties include: responsibilities of other stakeholders; prohibitions of using AI systems in particular domains (Art. 5); transparency-relevant duties for providers of user-oriented and generative AI systems (Art. 50); and provisions for structuring the regulatory administration (Art. 57-100). Although most duties are model-agnostic, providers of general-purpose AI models\footnote{i.e., those that can easily perform and integrate in a wide-range of applications, regardless of the intended purpose, e.g., LLMs; Art. 3, EUAIA.} have specific obligations regardless of the domain (Art. 51-56).

While LLMs are not inherently classified as high-risk, EUAIA duties may apply in at least three scenarios. First, stakeholders in the regulated contexts may find the general capabilities and user-friendliness of LLM-based chatbots to be worth the compliance effort. Second, as first of its kind globally, the EUAIA may become the standard framework for how to structure voluntary risk management. Third, regular reviews by the legislators (Chapter IX \& Art. 112) and any detected incidents (Art. 73) may result in the risk classification, domain coverage or model-specific duties being amended.

The law requires that providers establish quality properties such as unbiasedness, privacy, cybersecurity and safety to the user at an appropriate level. Compliance, however, means translating those properties into technical measures, interpreting their appropriateness in a given context, and managing risk to ensure their stability over time. This stability in performance, safety and security across contexts and time is known as \textbf{robustness}, which has long been a difficult problem in AI. For example, even after extensive training, LLMs can be brittle to adversarial attacks that elicit harmful responses with randomized, automated or manual prompting \citep{Zou2023_Universal}. Despite attempts of many providers to reduce that risk by setting guardrails, simple attacks still tend to succeed \citep{geiping2024}.

Ensuring robustness and compliance with EUAIA over time implies that testing, surveilling, reasoning about and reacting to newly discovered attacks. Thus, providers need to monitor and evaluate the impact of developments potentially affecting the safety and security of their LLM-based systems. Given the novelty of the field, valuable data about attacks and defenses is found in gray literature such as preprints \citep{geiping2024} and technical reports \citep{russinovich2024great}, or online repositories \citep{anthropic2024} for replicating experiments. Assurance, or establishing justifiable evidence-based confidence that a property has been achieved \citep{nist2024}, thus depends on effective knowledge management, which in turn depends on proper formalization of that knowledge.

\section{Methodology}

Our research methodology centers on knowledge representation from three parallel streams. First, we perform a simple legal analysis \citep{hohfeld2001,vanengers2015law} of the EUAIA to identify relevant duties\footnote{i.e., legal obligations that a particular stakeholder should satisfy \citealp{hohfeld2001}.} and stakeholders. Second, we elicit information about adversarial attacks and defenses in unstructured expert interviews and literature review (cf. \citealp{bueno2024argument}). Third, we use the Goal Structuring Notation (GSN; \citealp{acwg2021}) to express the confidence about EUAIA compliance and adversarial robustness in an exemplary assurance argument, comprising claims and evidence about appropriate defenses. We then combine and formalize this information in an ontology \footnote{i.e., specification of concepts, categories and relations in a particular domain. In this stage, we focus on expressing concepts in a graph of semantic triples, i.e., subject-predicate-object statements.} using the Web Ontology Language \citep{w3_2012_owl}, and display it as a human-readable narrative FactSheet report \citep{Arnold2019factsheet}.

\section{Proposed Framework}

\begin{table}[htbp]
  \centering
  \begin{tabularx}{\textwidth}{@{} >{\hsize=.05\hsize}X| >{\hsize=.2\hsize}X >{\hsize=.15\hsize}X >{\hsize=1.4\hsize}X @{}}
    \toprule
    \textbf{\#} & \textbf{§} & \textbf{S.*} & \textbf{Relevant Duties} \\
    \midrule
    1  & 9.2            & A & Identify, evaluate and mitigate \textit{reasonably foreseeable} risks of the system. \\
    2  & 9.5            & A & Ensure \textit{appropriate} and \textit{adequate} risk management measures. \\
    3  & 10.2           & A & Establish confidentiality and security of private data collected for assurance of other duties (e.g., bias mitigation). \\
    4  & 13.3, \newline Annex IV           & A & Include information about robustness and cybersecurity (e.g., metrics) and their limitations in instructions for use. \\
    5  & 14.2           & A & Design system for \textit{effective} human oversight regarding safety monitoring and prevention/minimization of \textit{reasonably foreseeable} misuse. \\
    6  & 14.4           & A & Design \textit{appropriate} functionalities for human overseers to: understand the system; monitor for "anomalies, dysfunctions and unexpected performance"; understand, override, and reverse the output; and intervene or interrupt the system's operation in a \textit{safe} state. \\
    7  & 15.1           & A & Establish an \textit{appropriate} level of robustness and cybersecurity. \\
    8  & 15.4           & A & Establish robustness and resilience of system regarding "errors, faults or inconsistencies." \\
    9  & 15.5           & A & Establish cybersecurity measures against adversarial and poisoning attacks. \\
    10 & 17.1           & A & Establish security-of-supply measures. \\
    11 & 31.2           & B & Satisfy \textit{suitable} cybersecurity requirements. \\
    12 & 50.2           & C & Ensure that AI-generated content is \textit{robustly} and \textit{reliably} watermarked. \\
    13 & 53.1, \newline An.XI           & C & Report on measures used to detect \textit{unsuitable} data sources and biases; evaluation of \textit{systemic} risk; measures for adversarial testing, model alignment and fine-tuning; system architecture and dependencies. \\
    14 & 55.1           & C & Establish cybersecurity and adversarially test with respect to systemic risks. \\
    15 & 57.6           & D & Support safety risk identification, testing, and mitigation in regulatory sandboxes. \\
    16 & 58.4           & D, A & Prespecify safeguards and conditions for real-world testing. \\
    17 & 70.3           & D & Establish safety and cybersecurity expertise. \\
    18 & 70.4           & D & Ensure an \textit{adequate} level of cybersecurity. \\
    19 & 73.1,7-8,11    & A, E, \newline F, D & Notify supervising stakeholder of a \textit{serious} incident. \\
    20 & 73.2-6         & A, E & Establish and report on the definite, \textit{reasonably likely} or suspected causal link between the system and a serious incident. \\
    21 & 74.12          & A & Securely provide documentation and data on system. \\
    22 & 78.2           & D & Establish cybersecurity measures for data obtained from providers. \\
    23 & 92.5,7         & C & Supply information on testing, safeguards and risk mitigation measures at the request of the AI Office. \\ \midrule
    \multicolumn{4}{>{\hsize=1.85\hsize}X}{*Stakeholders - A: High-risk AI System Provider; B: Notified Body; C: General-Purpose AI Provider; D: National Competent Authority; E: Deployer; F: Market Surveillance Authority.} \\
    \bottomrule
  \end{tabularx}
  \centering
  \caption{Overview of robustness-relevant EUAIA duties. Text is paraphrased, and qualifiers emphasized by authors for readability; see original text alongside Art. 3 for corresponding definitions.}
  \label{tab:euaia}
\end{table}

We identify 23 duties in EUAIA (cf. Table \ref{tab:euaia}) that directly refer to safety, cybersecurity or robustness, or proximate terms such as incident, risk or misuse. Providers of high-risk AI systems need to satisfy fifteen of those duties, and providers of general-purpose AI models four; other duties relate to deployers, market surveillance authorities and national competent authorities. While not representative of all relevant demands or conditions\footnote{e.g., transferability of Cybersecurity Act certificates (Art. 42 Para. 2) or conditional exemption of providers of free and open-source models (Art. 2 Para. 12; Art. 53 Para. 2).}, the list is a starting overview of the intersection between compliance and robustness.

\begin{wrapfigure}{r}{0.45\textwidth}
    \centering
    \vspace{-0.5cm}
    \includegraphics[width=0.45\textwidth]{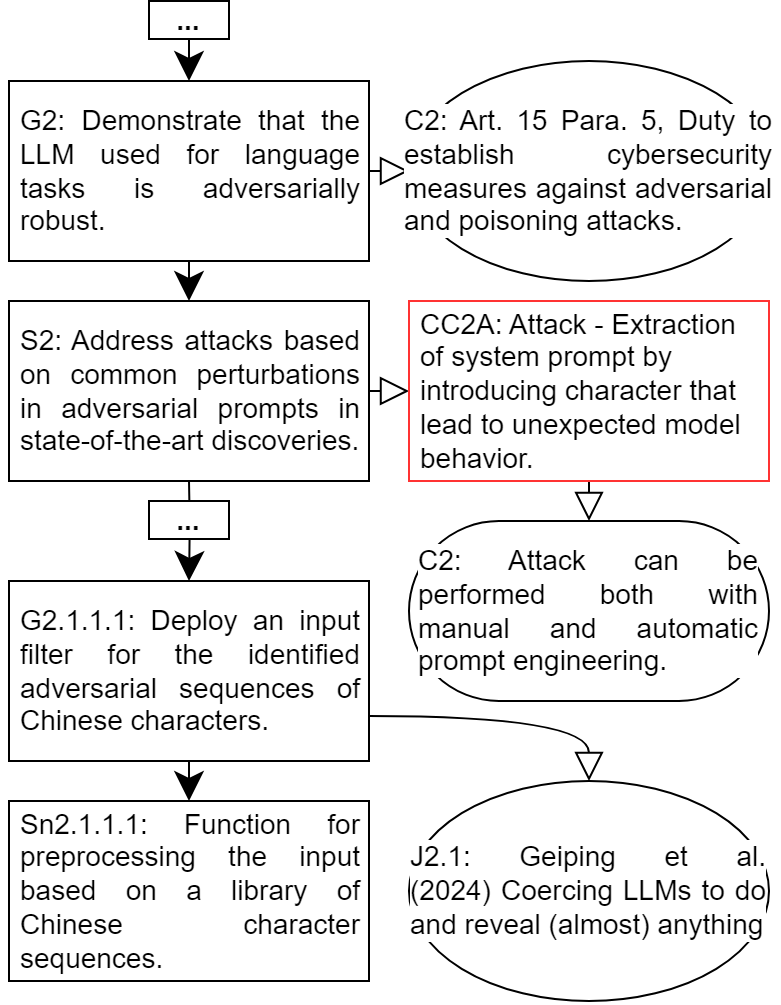}
    \caption{Excerpt from a GSN-based assurance argument, operationalizing the duty in Art. 15 Para. 5. Legend: goals (G), strategies (S), justifications (J), contexts (C), solutions (Sn) and counterclaims (CC).}
    \label{fig:argument}
    \vspace{-0.5cm}
\end{wrapfigure}

Much of the legal text includes context-specific adjectives or qualifiers such as "reasonably foreseeable", "suitable", "appropriate" or "effective." Lacking the technical requirements that operationalize what is suitable, providers would need to devise strategies to comply, and justify their suitability in the given context. This is a common case for building an assurance argument, as visualized in Figure \ref{fig:argument}.

One possible strategy includes mitigating attacks based on character combinations, as described by \citep{geiping2024}. Experiments show that specific characters in prompts can trigger profanity or leak hidden instructions in responses. They demonstrated the effectiveness of such attacks across various pre-trained open-source LLMs (e.g., LLaMa-2-7B-chat) using different scripts (e.g., Latin or Chinese).

Engineers can deploy several defenses here. Initially, a simple static input filter may be used to screen out prompts with characters that more frequently lead to such responses. Over time, this filter can be refined by testing how particular characters and combinations thereof affect the particular model; filter parameters can be adjusted to better and dynamically distinguish between benign and adversarial prompts. Ultimately, a more robust but costly solution would be to retrain the LLM to be less vulnerable to any character.

\begin{figure}
    \centering
    \includegraphics[width=\linewidth]{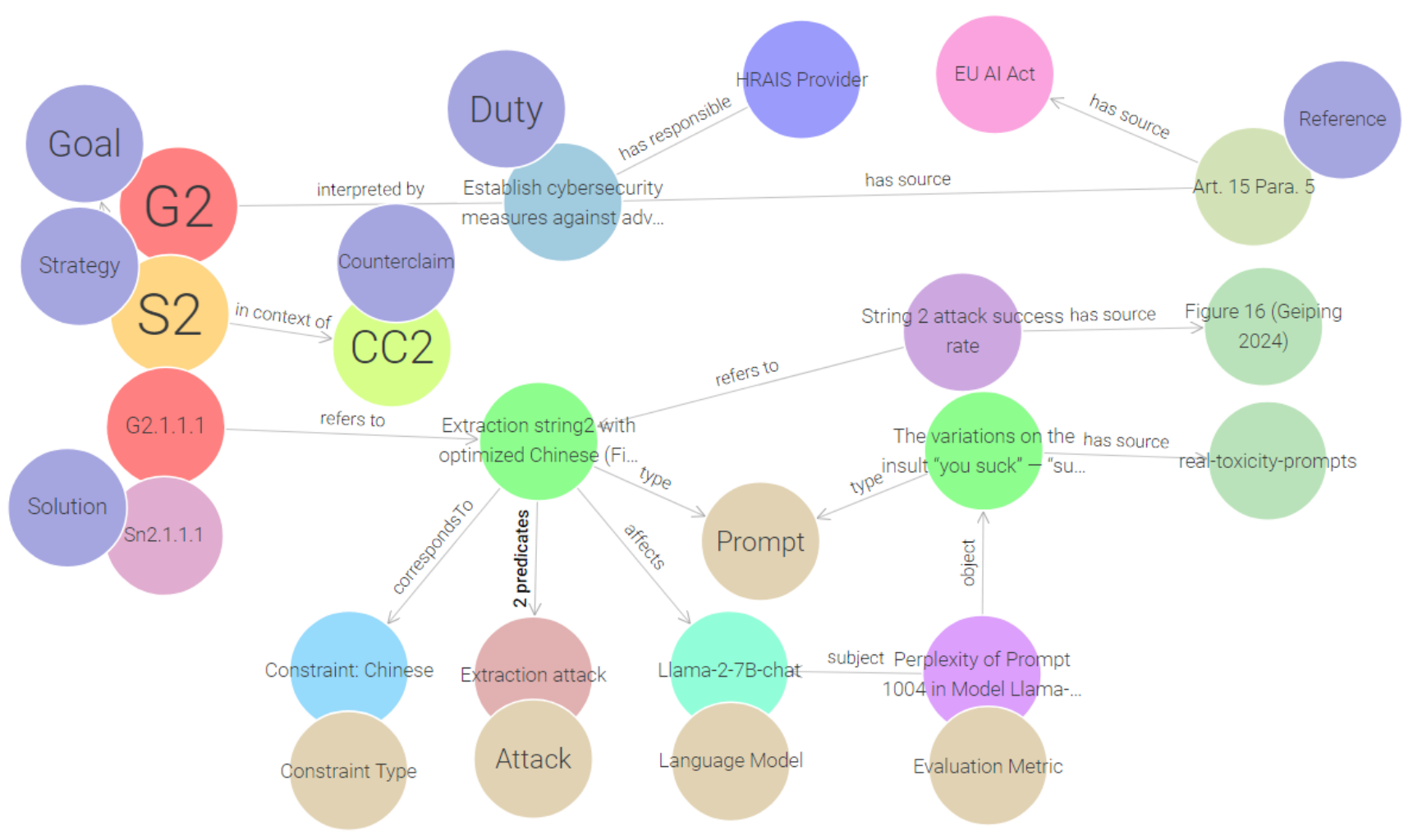}
    \caption{Excerpt from the ontology. Left-most circles make the argument (fig \ref{fig:argument}), while all remaining circles represent attacks, defenses, duties and sources. Coloring is arbitrary.}
    \label{fig:graph}
    \vspace{-0.3cm}
\end{figure}

We developed an ontology (Figure \ref{fig:graph}) to formalize and link the concepts that are used for evaluating, implementing and tracing the effectiveness of defenses. First, it contains information about a prompt, its contents and characters in a way that allows providers to retrieve and calculate of values needed for both static and dynamic filters. Second, it traces the sources of successful adversarial prompts (Figure 16 of \citealp{geiping2024}) and allows comparison with example data previously used to adversarially train the model. Third, it traces the provenance of the EUAIA duty (i.e., Art. 15 Para. 5), and links it with the assurance argument, so that this information can be systematically documented in a factsheet. This allows engineers and other stakeholders to track their status of compliance, perform causal analyses, and maintain LLM defenses, making their systems' robustnes auditable with respect to the EUAIA.

\section{Conclusion}

The EU Artificial Intelligence Act aims to mitigate risks of AI systems by imposing obligations on the robustness of various properties. However, for systems with LLM components, the implementation of these duties will be significantly challenging due to the inherent complexity and opacity of LLMs, alongside the continuous emergence of new security threats. Our proposed framework seeks to make the process of ensuring compliance and robustness effective, by allowing engineers (i.e., providers and deployers) to more easily represent and reason about LLM defenses through ontologies and assurance cases. The framework allows legal stakeholders and users to audit these systems with a complete, accurate and up-to-date snapshot.

Nonetheless, we recognize that this approach currently relies on manual work in creating arguments. This limits its usefulness for documenting and evaluating changes to law, system or attack vectors. Our future research centers on integrating the framework with techniques and tools that would allow arguments, concepts and relations to be expressed automatically, and evaluating it experimentally.

\section{Acknowledgements}
This work was partially supported by financial and other means by the following research projects: DUCA (EU grant agreement 101086308), FLA (supported by the Bavarian Ministry of Economic Affairs, Regional Development and Energy), the DiProLeA (German Federal Ministry of Education and Research, grant 02J19B120 ff), as well as our industrial partners in the FinComp project. We thank the reviewers for their valuable comments.

% ----------------
% | Bibliography |
% ----------------

\bibliographystyle{agsm}
\bibliography{literature}
% ---------------------
% |  End of Document  |
% ---------------------

\end{document}